\documentclass[aps,prl,twocolumn,showpacs,groupedaddress,amsmath,amssymb]{revtex4}

\usepackage[pdftex]{graphicx}
\usepackage{color}
\usepackage{textcomp}

\renewcommand{\section}[1]{\textit{#1} --}

\begin{document}

\title{Random Matrix Theory approach to Mesoscopic Fluctuations of Heat Current}

\author{Martin Schmidt$^{1}$, Tsampikos Kottos$^{1,2}$}
\affiliation{$^1$Department of Physics, Wesleyan University, Middletown, Connecticut 06459}
\affiliation{$^2$Max Planck Institute for Dynamics and Self-organization (MPIDS),  37077 G\"{o}ttingen, Germany}
\author{Boris Shapiro}
\affiliation{Technion - Israel Institute of Technology, Technion City, Haifa 32000, Israel}
\begin{abstract}
We consider an ensemble of fully connected networks of $N$ oscillators coupled harmonically with random springs and show, using Random 
Matrix Theory considerations, that both the average phonon heat current and its variance are scale-invariant and take universal values in the 
large $N$-limit. These anomalous mesoscopic fluctuations is the hallmark of strong correlations between normal modes.
\end{abstract}
\maketitle
{\it Introduction--} 
The study of heat conduction by phonons in disordered or chaotic structures have attracted recently considerable interest \cite{LLP03,D08,
LXXZL12}. A central issue of these investigations is the dependence of the average heat current $J$ on the system size $N$. A naive 
expectation is that disorder or phonon-phonon interactions scatters normal modes and induces a diffusive energy transport that leads 
to a normal heat conduction described by Fourier's law which states that $J\sim N^{-1}$. Many studies \cite{D08,LXXZL12,LZH01,DL08,
LD05,RD08,KCRDLS10}, however, find that in low dimensional chains $J$ scales as $J \sim N^{-\alpha}$, where $\alpha$ is usually different 
from one. In fact, experiments on heat conduction in nanotubes and graphene flakes have reported observations of such anomalous behavior 
\cite{COGMZ08,NGPB09,LRWZHL12}.

However, many real stuctures such as biological systems \cite{D98} and artificial networks in thin-film transistors and nanosensors \cite{HHG04}
are not simple one-dimensional or two-dimensional lattices. Rather they are characterized by a complex connectivity that can be easily 
designed and realized in the laboratory \cite{KMA05,COMZ06,PCWGD05} . Therefore, not only is it a fundamental demand for the development 
of statistical physics to understand normal and anomalous heat conduction in complex networks of coupled oscillators, but it is also of great 
interest from the technological point of view, since the achievements of modern nano-fabrication technology allow us to access and utilize 
such structures with sizes in the range of a few nanometers up to few hundred nanometers.  

The complexity of coherent wave interferences in such networks calls for a statistical treatment of any of their transport characteristics. This 
way of thinking has been adopted already in classical wave and quantum transport theories associated with mesoscopic chaotic or disordered 
systems and resulted in a plethora of exciting results \cite{S99,B97,A00} like the weak and strong Anderson localization, the universal conductance 
fluctuations (UCF) etc. The statistical approach led also to the revival of Random Matrix Theory (RMT) \cite{W57}, as a major theoretical tool 
for the analysis of transport characteristics of complex systems. RMT has found applications in many areas of physics ranging from nuclear, 
atomic and molecular physics to mathematical physics (for a review see \cite{Oxford}). Consequently a variety of RMT ensembles have been 
introduced \cite{Z96}, extending the original work of Wigner beyond the traditional Gaussian ensembles, helping to understand phenomena 
like the Quantum Hall effect \cite{H95}, Anderson localization and Metal-to-Insulator transition \cite{EM08}. The success of RMT 
was such that in recent days it has become almost a dogma that this theory captures the universal properties of complex disordered and 
chaotic systems \cite{S99,A00,B97,Oxford}. It is thus surprising, that the study of fluctuations and the use of RMT as a concrete tool for their 
analysis were not brough up in any of the previous studies of heat transport.
 
In this Letter we address heat transport and the associated sample-to-sample mesoscopic fluctuations of complex networks of $N$ equal 
masses connected  with one another via random harmonic springs. The force matrix that describes the dynamics of the system is real symmetric 
and consists of random elements (spring constants). We find that the statistical description of heat transport can be effectively described by 
an ensemble of Random Matrices with diagonal elements that fluctuate with a variance $N$ times larger than the corresponding variance of 
the off-diagonal elements. Using RMT considerations we show that both the average heat current $\langle J\rangle$ and its variance $(\Delta 
J)^2$ are scale-invariant and get a universal value in the large-$N$ limit. These anomalous mesoscopic fluctuations is the hallmark of strong 
correlations between normal modes of the system. For moderate size networks, with random springs taken from a  distribution with 
variance $\sigma^2 < 1/N$, we find that the heat  transport is sensitive to the boundary conditions imposed on the two  end-sites which 
are coupled to the thermal baths. In  particular, for fixed boundary conditions the current is completely dominated by a pair of surface 
modes, for which only the end sites oscillate with appreciable amplitude. We hope that our analysis will motivate the use of RMT models and 
provide new insight in the mesoscopic fluctuations of heat transport.


\section{Fully Connected Harmonic Networks } We consider a network of $N$ harmonic oscillators of equal masses $m=m_0$. 
The system is described by the Hamiltonian  \cite{note1}
\begin{equation}
{\cal H}={1\over 2} P^T {\hat M}^{-1} P + {1\over 2} Q^T {\hat \Phi} Q
\label{eq:H} 
\end{equation}
where $Q^T\equiv(q_1,q_2,\cdots,q_N)$, $P^T\equiv(p_1,p_2,\cdots,p_N)$ and $q_n,p_n$ are respectively the individual oscillator displacements 
and momenta. The mass matrix is $M_{nm}=\delta_{nm} m_0$, and ${\hat \Phi}$ is the force matrix that contains also information about the boundary 
conditions (b.c.). For a fully connected network of coupled oscillators with free b.c. ${\hat \Phi}$ takes the form $\Phi_{nm}=
(\sum_l k_{nl})\delta_{nm} - k_{nm}$ where $k_{nm}$ are the spring coupling constants. These spring constants $k_{nm}$ are chosen to be 
symmetric ($k_{nm}=k_{mn}$) and uniformly distributed according to $k_{nm} \in \left[-\frac{W}{2}+1, \frac{W}{2}+1\right]$ where the disorder 
strength parameter $W$ has to be smaller than $2$ in order to ensure that $k_{nm}\geq 0$.  In the case of fixed b.c. ${\hat \Phi}$ has to be modified 
by considering the coupling of the first and last oscillator to hard walls i.e. $\Phi_{nm}^{\rm fix}=\Phi_{nm}+(k_{01}\delta_{n1}\delta_{m1}+k_{NN+1}
\delta_{nN}\delta_{mN})$.  

Next, we want to study the non-equilibrium steady state (NESS) of this network driven by a pair of Langevin reservoirs set at temperatures $T_1^{\rm 
B}$ and $T_N^{\rm B}$ (we assume $T_1^{\rm B}>T_N^{\rm B}$), and coupled to the first $n=1$ and last $n=N$ masses with a constant coupling 
strength $\gamma$. The corresponding equations of motion that describe also the coupling to the bath are  $\dot{q}_n=\partial {\cal H}/\partial 
p_n, \;  \dot{p}_n=-\partial {\cal H}/\partial q_n+\left(-\gamma p_n/m_0 + \sqrt{2\gamma T_n^{\rm B}} \zeta_n\right) (\delta_{n1}+\delta_{nN})$, 
where $\zeta_n(t)$ is delta-correlated white noise $\overline{\zeta_n(t) \zeta_{n'}(t')} = \delta_{nn'} \delta(t-t')$. The NESS current is evaluated as 
$\overline{J}={\gamma\over m_0} (T_1^{\rm B}-T_1)={\gamma\over m_0} (T_N-T_N^{\rm B})$ where the temperature of the $n$-th oscillator is 
defined as $T_n\equiv\overline{ p_n^2}/m_0$. The notation $\overline{\cdots}$ which will be implicetly assumed from now on, indicates the 
thermal statistical average. 

For weak coupling $\gamma$ it was shown in Ref.~\cite{D08,LLP03} that
\begin{equation}
\label{current1}
J=\sum_{\mu} J^{(\mu)};\quad J^{(\mu)}= C_0 {I_1^{(\mu)}I_N^{(\mu)} \over I_1^{(\mu)}+ I_N^{(\mu)}}
\end{equation}
where $I_n^{(\mu)}\equiv |\psi_n^{(\mu)}|^2$ and $\psi_n^{(\mu)}$ indicates the $n-$th component of the $\mu$-th normal mode of the Hamiltonian
Eq. (\ref{eq:H}) and the coefficient $C_0\equiv {\gamma\over m_0} (T_1^{\rm B}-T_2^{\rm B})$ \cite{note0}. In Eq. (\ref{current1}), the $\mu$-th addendum 
$J^{(\mu)}$ is naturally interpreted as the contribution of the $\mu$-th mode to the
total heat flux $J$. As intuitively expected, $J^{(\mu)}$ is larger for modes that have larger amplitudes at the boundaries and couple thus more strongly 
with the reservoirs. Thus the analysis of heat flux $J$ reduces to the study of the normal modes of Hamiltonian ${\cal H}$ given by Eq. (\ref{eq:H}).

\begin{figure}
   \includegraphics[width=1\linewidth, angle=0]{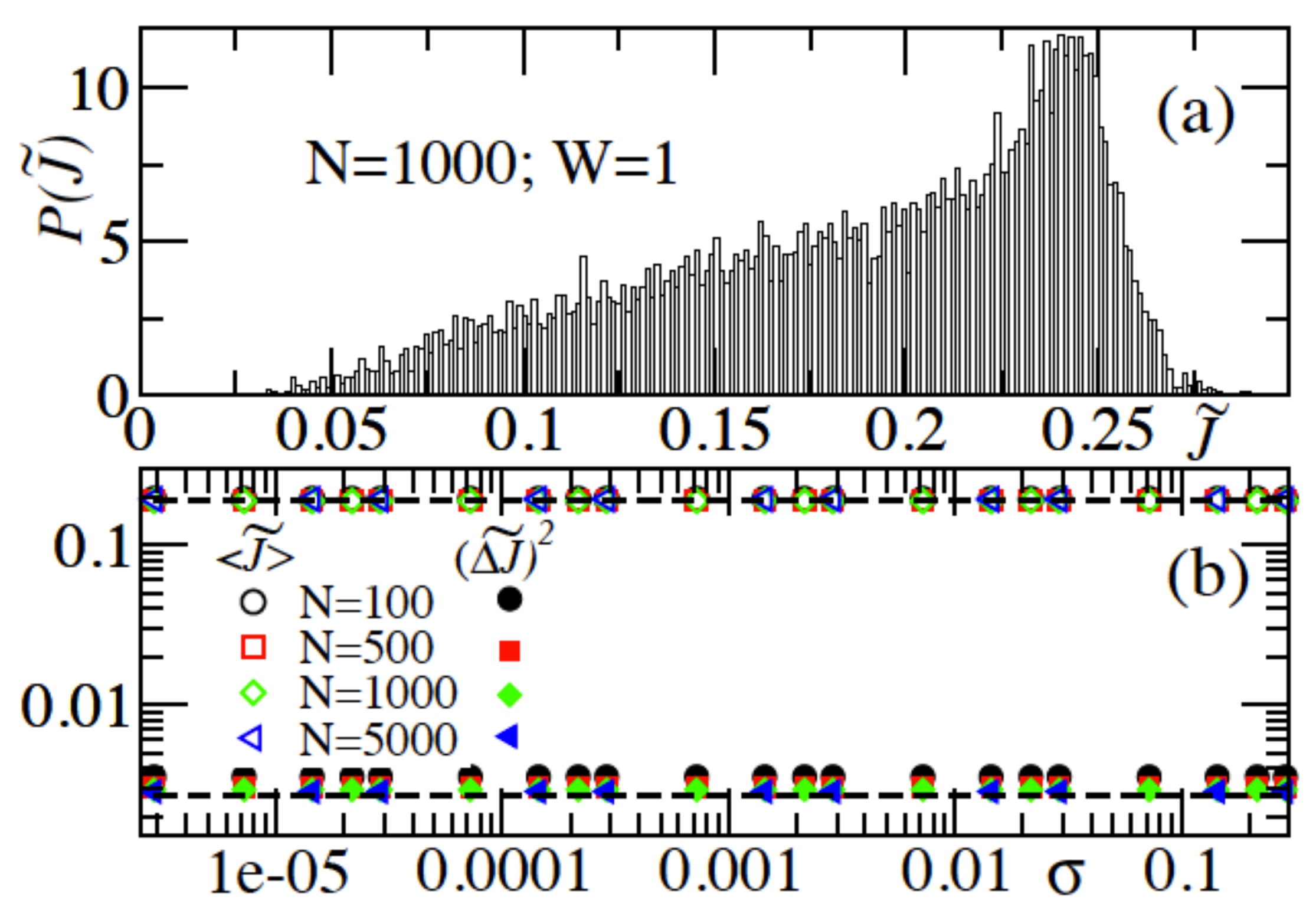}
    \caption{(Color online) Free boundary conditions. (a) A typical distribution of rescaled heat flux ${\tilde J}\equiv  J/C_0$ for a network of $N=10^3$ 
oscillators and disorder strength $W=1$.  (b) The rescaled average heat current $\langle {\tilde J}\rangle$ (open symbols) and 
variance $\widetilde{(\Delta J)^2}\equiv (\Delta J)^2/C_0^2$ (filled symbols) versus $\sigma$. Various 
system sizes $N$ (indicated in the figure) have been used. The dashed lines in (b) are the results of $D$-RMT ensemble with strongly 
fluctuating diagonal elements. Both in (a) and (b) we have used Eq. (\ref{current1}) in order to evaluate the heat current $J$.
\label{fig1}
}
\end{figure}

\section{Random Matrix Theory Formulation } We separate out the random component of the spring constants and re-write them as $k_{nm}=
1+W_{nm}$ where now $W_{nm} \in \left[-\frac{W}{2}, \frac{W}{2}\right]$. The force matrix ${\hat \Phi}$ can be decomposed into a constant matrix 
${\hat A}$ and a random part ${\hat R}$ as
\begin{equation}
{\hat \Phi}={\hat A} + {\hat R} \quad {\rm where} \quad {\hat A}\equiv N {\hat 1} - {\hat U}; \quad {\hat R}\equiv {\hat D} - {\hat W}
\label{phi}
\end{equation}
where ${\hat 1}$ is the $N\times N$ unit matrix, ${\hat U}$ is a matrix whose all elements are equal to unity i.e. $U_{nm}=1$, ${\hat D}$ is a 
diagonal matrix with $D_{nn}=-\sum_{l\neq n} W_{nl}$, and ${\hat W}$ is a random matrix defined below. The above decomposition 
allow us to distinguish the various contributions. The matrix ${\hat W}$ can be treated as a "standard"
RMT ensemble (note though that it has zero diagonal elements). It is convenient to rewrite it as ${\hat W}=\sigma {\hat W_0}$ where ${\hat W_0}$ is a RM with elements having unit variance 
where $\sigma^2\equiv (\Delta W_{nm})^2=W^2/12$. The diagonal matrix ${\hat D}$ has Gaussian distributed random elements with 
$\langle D_{nn}\rangle=0$ and variance $(\Delta D_{nn})^2=(N-1) \sigma^2$. 

The constant matrix ${\hat A}$ can be diagonalized exactly. It has (a) one eigenvalue $\omega_0=0$ with a corresponding eigenvector
$(1/\sqrt{N}) (1,1,1\cdots,1)^T$ and (b) $N-1$ degenerate eigenvalues $\omega_{\mu}=N$ ($\mu=1,2,\cdots N-1$). Now consider adding
to ${\hat A}$ the random matrix ${\hat W}$, i.e. we neglect for the moment ${\hat D}$ and consider 
\begin{equation}
\label{phi_prime}
{\hat \Phi}'= {\hat A} + \sigma {\hat W_0}. 
\end{equation}
Already for an arbitrary small 
$\sigma$, the $N-1$ degeneracy will be removed and the corresponding eigenvectors will be those  of a $(N-1) \times (N-1)$ random 
matrix. The $(N-1)$-time degenerate level is broadened into a band of width $\sim \sigma \sqrt{N}$. The perturbation theory applies for 
$\sigma \sqrt{N} < N$ i.e. for $\sigma < \sqrt{N}$. However even
for larger $\sigma$ the RMT still applies because then we can simply neglect the matrix ${\hat A}$ in Eq. (\ref{phi_prime}). In short, for 
small $\sigma$ we have an RMT for $(N-1)$-rank matrices (the contribution to current of the level with $\omega_0\approx 0$ can
be neglected, in comparison to the $(N-1)$-levels), whereas for large $\sigma$ we have an RMT for $N$-rank matrices. Thus, in the
large $N$ limit we treat Eq. (\ref{phi_prime}) as an ensemble of $N\times N$ GOE matrices \cite{S99}.

Normalization requires that $\langle I_1^{(\mu)}\rangle = \langle I_N^{(\mu)}\rangle =\langle I_n\rangle= 
1/N$. Defining a rescaled variable $X_n^{(\mu)}=I_n^{(\mu)}/\langle I_n \rangle$, we can rewrite Eq. (\ref{current1}) as
\begin{equation}
\label{current2}
J= {C_0\over N} Z;\quad Z=\sum_{\mu=1}^N z^{(\mu)}; \,\, z^{(\mu)}\equiv  {X_1^{(\mu)}X_N^{(\mu)} 
\over X_1^{(\mu)}+ X_N^{(\mu)}}
\end{equation}
According to the standard RMT, and omitting the mode label $\mu$, the joint probability distribution of the rescaled eigenmode intensities 
$X_n$ is a product of two Porter-Thomas distributions $P(X_1,X_N)= (1/2\pi) (1/\sqrt{X_1X_N}) \exp[-(X_1+X_N)/2]$. Assuming further 
that the various $z^{(\mu)}$-terms appearing in Eq. (\ref{current2}) are statistically independent we get 
\begin{equation}
\label{current_RMT}
\langle J\rangle = {1\over 4} C_0;\,\,  (\Delta J)^2 ={1\over 8N} C_0^2 
\end{equation}
Comparison of these theoretical predictions with a direct numerical evaluation of the mean and the variance of heat current $J$ 
via Eq. (\ref{current1}) (see Fig. \ref{fig1}b) leads us to conclude that standard RMT considerations describe well the scaling of 
the {\it average} current but {\it not the variance}. To obtain the correct description of the variance, it is necessary to treat the 
full force  matrix, as given in Eq. (\ref{phi}). Below we show that the matrix $D$ induces strong correlations between different 
$z^{(\mu)}$'s, thus, invalidating the assumption which led to Eq. (\ref{current_RMT}) for the variance \cite{note2}. 

\section{D-RMT ensemble with strongly fluctuating diagonal elements } We now consider the ensemble of matrices given by Eq. (\ref{phi}). 
Again for large $N$, the matrix ${\hat A}$ has no effect, so it is enough to understand the eigenvectors of the random matrix ${\hat R}$.
The eigenvalues of ${\hat D}$ are of order $|D_{nn}|\sim \sigma \sqrt{N}$, so that they occupy a band of order $\sigma \sqrt{N}$ and 
are separated by a typical energy interval $\sigma \sqrt{N}/N = \sigma/\sqrt{N}$. The same is true for the eigenvalues of the matrix 
${\hat W}$. In this sense ${\hat D}$ and ${\hat W}$ are ``of the same strength" and neither can be treated as perturbation to the other. 
However, the qualitative understanding of the eigenvectors of the combined matrix ${\hat R}$ is along the following lines: The eigenvectors 
of ${\hat D}$ are localized on the individual sites i.e. the $\mu-$th eigenvector is $\psi_n^{(\mu)}=\delta_{n\mu}$. 
The matrix ${\hat W}$ mixes these eigenvectors, so that eigenvectors of ${\hat R}$ are spread over all sites and resembles those of a standard
RMT. Therefore $\langle J\rangle$ is qualitatively not different from the standard RMT result of Eq. (\ref{current_RMT}). The only difference 
is that the coefficient $1/4$ now assumes the numerical value $\approx 0.19$ (see Fig. \ref{fig1}b). 

As far as the variance $(\Delta J)^2$ is concerned we get results that are qualitatively different from the standard RMT result 
of Eq. (\ref{current_RMT}). It turns out that in this case  each eigenvector of ${\hat R}$ "remembers" the set $(D_{11},\cdots,D_{NN})$ 
of the eigenvalues of ${\hat D}$ so that correlations between different eigenvectors of ${\hat R}$ are significantly stronger than those 
for the standard RMT. Namely the mode-mode correlations between the different $z^{(\mu)}$'s of the matrix ${\hat R}$ 
are described by \cite{note2}
\begin{equation}
\label{correlations}
\langle z^{(\mu)} z^{(\nu)}\rangle = \langle z^{(\mu)} \rangle \langle z^{(\nu)} \rangle (1+\epsilon) =  \langle z \rangle^2 (1+\epsilon)
\end{equation}
where $\epsilon$ is a constant. Using Eq. (\ref{correlations}) we calculate the variance $(\Delta Z)^2$ of the random variable 
$Z$ (see Eq. (\ref{current2})):
\begin{equation}
\label{var_f_matrix}
(\Delta Z)^2 = N^2 \epsilon \langle z^2\rangle + O(N).
\end{equation}
Expressing $(\Delta J)^2$ in terms of $Z$ via Eq. (\ref{current2}) we get
\begin{equation}
\label{var_ful}
(\Delta J)^2 = C_0^2 \epsilon \langle z\rangle^2,\quad \langle z\rangle \approx 0.19
\end{equation}
Direct numerical evaluation of the variance $(\Delta J)^2$ based on Eq. (\ref{current1}) confirms the above theoretical estimates. In 
Fig. \ref{fig1}b we show some of our numerical results for rescaled variance $\widetilde{(\Delta J)^2}\equiv(\Delta J)^2
/C_0^2$. The data clearly indicate that $(\Delta J)^2$ is scale invariant for any disorder strength $\sigma$. Further $1/N$ numerical 
analysis allow us to extract the asymptotic value $\epsilon \approx 0.075$.

We have also checked that correlations between the matrices ${\hat D}$ and ${\hat W}$ do not play a role in our arguments. Detail 
numerical analysis indicates that if instead of the actual ${\hat D}$ (i.e. $D_{nn}=-\sum_i W_{ni}$) we consider a diagonal random 
matrix completely independent of ${\hat W}$ so that $(\Delta R_{nm})^2=\sigma^2 [1+(N-1)\delta_{nm}]$, we still obtain the 
same behavior for $\langle J\rangle$ and $(\Delta J)^2$ (dashed lines in Fig. \ref{fig1}b). We remark that this kind of ensembles, 
with strongly fluctuating diagonal elements, ($D$-RMT ensembles) have previously appeared in the context of mescoscopic physics 
\cite{S94}. 

\section {Fixed b.c.} Finally we investigate the effect of b.c. on the statistics of heat flux. We consider the other limiting case of 
fixed b.c. We assume that the first and the last oscillator are coupled to the left and right walls with spring constants $k_{01}=
1+W_{01}$ and $k_{NN+1}=1+W_{NN+1}$ respectively which are taken from the same ensemble of random springs as the ones 
in the bulk of the network. The random components are then included in the matrix elements $D_{11}$ and $D_{NN}$, respectively. 
The constant matrix ${\hat A}$ also changes to ${\hat A}^{\rm fix}={\hat A}+{\hat C}$ where $C_{nm}=\delta_{n1} \delta_{m1}+
\delta_{nN}\delta_{mN}$. This results in a  slight shift of the zero mode $\omega_0=0$ of the matrix ${\hat A}$ together with a 
``deformation" of the  $(1,1\cdots,1)^T$ eigenvector. Contribution of this level to the total current is of order $1/N$, and it is 
disregarded below.

\begin{figure}
   \includegraphics[width=1\linewidth, angle=0]{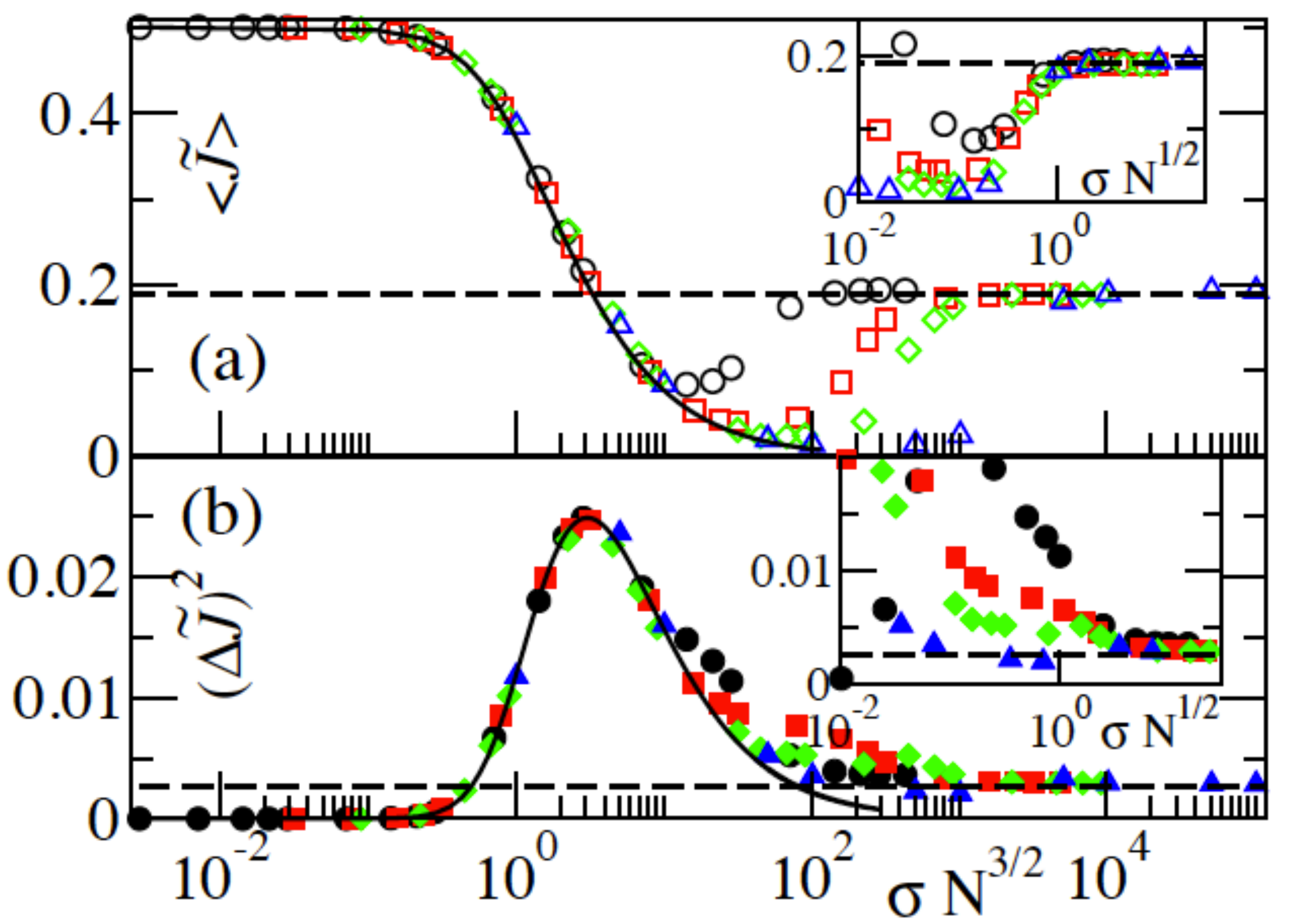}
    \caption{(Color online) Fixed boundary conditions. (a) The rescaled average heat current $\langle{\tilde J}\rangle\equiv \langle J
\rangle/C_0$ versus disorder strength for various system sizes $N$. In the main panel we scale the $x$-axis as $\sigma N^{3/2}$ 
while in the inset (RMT domain) we scale it as $\sigma N^{1/2}$. (b) The same as in subfigure (a) but now for the rescaled variance 
$(\widetilde{\Delta J})^2\equiv (\Delta J)^2/C_0^2$. The various symbols correspond to different system sizes $N$ as indicated in Fig. 
\ref{fig1}b. The dashed lines are the predictions of D-RMT (see also Fig.\ref{fig1}b) while the solid lines represent the theoretical predictions 
of Eq. (\ref{av2},\ref{var2}).
}
\label{fig2}
\end{figure}

In addition two new levels emerge from  the $N-1$ degenerate subspace of the matrix ${\hat A}$. The first one has the highest energy 
$\omega_{N-1}=N+1$ with a corresponding eigenmode $\psi^{(N-1)}=(1/\sqrt{2})(1,0,\cdots,-1)^T$. This is an exact eigenvalue 
and eigenvector of ${\hat A}^{\rm fix}$. The second level is slightly lower than $N+1$ (approximately by $2/N$) and its eigenvector 
is symmetric i.e. $\omega_{N-2}\approx N+1-2/N$ with $\psi^{N-2}\approx (1/\sqrt{2}) (1-1/N, -2/N,\cdots,-2/N, 1-1/N)^T$. 
Below we refer to these states as ``surface" modes.

It turns out that for a network described by the constant force matrix ${\hat A}^{\rm fix}$, most of the current is carried by the two
surface modes. Using Eq. (\ref{current1}) we find that $J^{(N-1)} = J^{(N-2)}= {1\over 4}C_0$. 
At the same time, the remaining $N-3$ degenerate modes {\it does not contribute} to 
the current (in the large $N-$limit). Since  any of these eigenvectors $\psi^{(\mu)}$ 
has to be orthogonal to both $\psi^{(N-1)}$ and $\psi^{(N-2)}$ we get that $\psi_1^{(\mu)}=\psi_N^{(\mu)}$ and 
$\psi_1^{(\mu)}=-\psi_N^{(\mu)}$. These two constrains are satisfied simultaneously only if $\psi_1^{(\mu)}=
\psi_N^{(\mu)}=0$ for any $\mu=1,\cdots,N-3$. Thus the total heat current is
\begin{equation}
\label{fix_aver_J}
J=\sum_{\mu=1}^N J^{(\mu)} \approx  {1\over 2} C_0
\end{equation}
The above result will still hold as long as the random matrix ${\hat R}$ does not destroy the pair of states $\psi^{(N-1)}$ and 
$\psi^{(N-2)}$. As $\sigma$ increases we observe a coupling of the two states towards a linear combination i.e. $(1/\sqrt{2})
[\psi^{(N-1)}\pm\psi^{(N-2)}]$. The origin of this re-organization is traced 
to the matrix ${\hat D}$ which in the $\{\psi^{(N-1)},\psi^{(N-2)}\}$ subspace, would produce a pair of eigenvalues 
separated by a distance of order $\sigma \sqrt{N}$. This has to be compared to the separation of order $1/N$ between the surface 
mode eigenvalues $\omega_{N-1}$ and $\omega_{N-2}$ of the matrix ${\hat A}^{\rm fix}$. When $\sigma$ reaches a value $\sigma_c
\sim N^{-3/2}$ the two "surface" eigenstates are destroyed giving rise to a set of new modes that have components $(0,\cdots,0,1)^T$ 
and $(1,0,\cdots,0)^T$ i.e. they are localized asymmetrically at the reservoir sites.  Consequently, the average current will drop to 
approximately a zero value. As the disorder continues to increase, the matrix ${\hat W}$ lifts the degeneracy of the $N-3$ levels 
centered around $\omega=N$ and creates a spectral band of size $\delta_W\sim \sigma \sqrt{N}$. For some critical value of $\sigma=
\sigma_{\rm RMT}\sim1/\sqrt{N}$ the bandwidth $\delta_W$ becomes as broad as the gap that separates the degenerate states from the 
surface states. The latter now merge with the continuum of states in the band, and the RMT results are recovered.

For disorder strength such that the dominant contribution comes only from the two surface states, a quantitative description of the 
heat transport can be achieved by considering a simple two level system. The two surface states of the perfect system are described 
(in the site representation) by the $2\times 2$ matrix ${\hat A}^{(2)} = -{1\over N} {\hat U}^{(2)}$ where $U^{(2)}$ has unit elements 
$U^{(2)}_{nm}=1$. This matrix has eigenvalues $\omega_1=0, \omega_2=-2/N$ and corresponding eigenvectors $\psi^{(1)}=(1/\sqrt{2}) 
(1, -1)^T$ and $\psi^{(2)}=(1/\sqrt{2}) (1,1)^T$. The two energy levels are separated by an interval $2/N$ where we have set the 
energy of the highest level (associated to the the $\omega_{N-1}$ level of the original problem) to zero. We now add the diagonal matrix 
${\hat D}^{(2)}$ with elements $D^{(2)}_{nm}=\sqrt{N} W_{nm}\delta_{nm}$ where $W_{nm}\in [-W/2,W/2]$. The total ``Hamiltonian" takes 
the form ${\hat \Phi}^{(2)} = {\hat A}^{(2)} + {\hat D}^{(2)}$. We can diagonalize exactly this two-dimensional matrix and get the 
corresponding eigenvectors. Using Eq. (\ref{current1}) we obtain $ J_2 = {2 C_0\over 4 + N^3 ({ W_{11}-W_{22}})^2}$.  From this we 
can further calculate the average and the variance of heat current. We get
\begin{equation}
\label{av2}
\langle J_2\rangle = 2 C_0{w\arctan \left[ {w\over 2}  \right] - \log\left[1+({w\over 2})^2\right]
\over w^2}; \quad w=N^{3/2} W
\end{equation}
while for the variance we get 
\begin{equation}
\label{var2} 
(\Delta J_2)^2={w^3 \arctan\left[{w\over 2} \right] - 8 \left(\log\left[1 + ({w\over 2})^2\right]
-w\arctan\left[{w\over 2} \right]  \right)^2 
\over 2 w^4 }C_0^2
\end{equation}
These theoretical predictions are compared in Fig. \ref{fig2} with the numerically evaluated average heat current and variance via 
Eq. (\ref{current1}) for various system sizes $N$ and disorder strength $W$. Obviously Eqs. (\ref{av2},\ref{var2}) do not apply for
$\sigma_{\rm RMT}\geq N^{-1/2}$ when RMT dominates the transport.

\section{Conclusions} In conclusion, we have employed RMT modeling as a valuable tool for the analysis of mesoscopic fluctuations 
of heat current $J$ in complex (chaotic) networks. For the most basic chaotic system consisting of a fully connected network of 
random springs we have found that both the average heat current and its variance are scale-invariant. For large $N-$limit, these 
quantities assume a universal value which is independent of the specific boundary conditions. Our analysis indicated that the statistical 
properties of $J$ are affected by the existence of correlations between normal modes. For moderate size networks with random springs 
taken from a distribution with variance $\sigma^2< 1/N$, the mean and the variance of heat current are affected by the existence 
of two surface modes emerging in the presence of fixed boundary conditions. It would be interesting to investigate the statistical 
properties of heat current for other geometries beyond the zero-dimensions, or in the presence of anharmonicities \cite{LLP03,GN94} 
and establish analogies with mesoscopic phenomena observed in the realm of electron transport.

\acknowledgments{This research was supported by an AFOSR No. FA 9550-10-1-0433 grant, and by the DFG Forschergruppe 760. 
(TK) acknowledge T. Prosen for useful discussions. (B.S), thanks the Wesleyan Physics Department for hospitality extended to
him during his stay, when the present work had been done.}

\begin{thebibliography}{99}
\bibitem{note1} Notice that this is a ‘‘scalar’’ phonon model, where the vectorial properties of the
modes have not been taken into consideration (and thus the
matrix has $N$ and not $3N$ modes).
\bibitem{note2} In the standard RMT case different eigenvectors  
(and, thus, different $z^{(\mu)}$'s) are only weakly correlated, due to  
their orthogonality, i.e. $\langle z^{(\mu)} z^{(\nu)}\rangle = \langle z^{(\mu)} \rangle \langle z^{(\nu)} \rangle + {a\over N} =  
\langle z \rangle^2 + {a\over N}$ with some constant coefficient $a$.   
These weak correlations result in the replacement of the coefficient  
$1/8$ in Eq. (\ref{current_RMT}) by the (numerically evaluated) coefficient $1/16$.
\end{thebibliography}

\newpage

{\bf Supplementary Material: NESS for the fully Connected Network}

\begin{figure}
   \includegraphics[width=1\linewidth, angle=0]{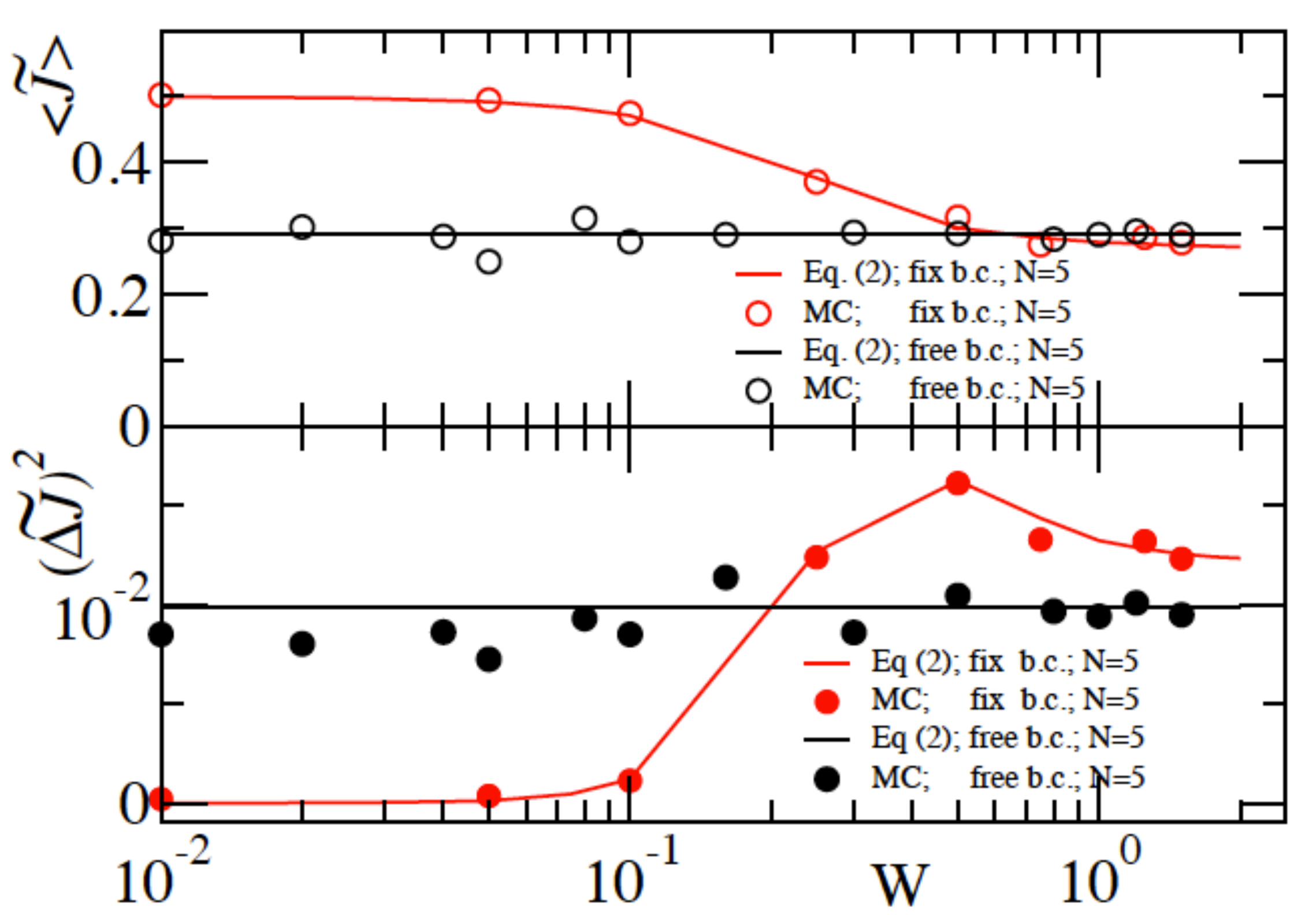}
    \caption{(Color online) Molecular Dynamics (MD) simulations (open and filled symbols) for the case of a network of $N=5$ fully connected 
oscillators. The results from the MD are compared with the results coming from Eq. (\ref{current1}). A nice agreement, both for the mean 
heat current (upper) and the variance (lower) is observed, indicating that our system can reach a NESS.
\label{fig3}}
\end{figure}

In order to establish that the fully connected network
of harmonic oscillators Eq. (\ref{eq:H}) reaches the NESS, we have also performed independent Molecural Dynamics (MD) simulations 
for both free and fixed boundary conditions. Since these simulations are time consuming we confine ourselves to moderate $N$-sizes. 
In Fig. \ref{fig3}  we repost such representative simulations for a case of a fully connected network of 
$N=5$ coupled oscillators with random springs $k_{nm}$ taken from a uniform distribution $k_{nm}\in [1-W/2;1+W/2]$ and 
compare these results with the ones coming from a direct diagonalization of the associated force matrix ${\hat \Phi}$
with the use of Eq. (\ref{current1}). 

In Fig. \ref{fig3}  open symbols correspond to the average heat current and full symbols to its variance evaluated from the MD 
simulations, while the solid lines are the results of the diagonalization method that makes use of Eq. (\ref{current1}). For the MD 
simulations we have used typically 100 disorder realizations (this has to be compared to the diagonalization method where 
typically we had more than $5000$ realizations). An additional time average (over the last $20$ time units) was performed in 
order to average out the oscillations of the chain elements. In order to check the convergence of the MD simulations, we have 
compared the flux $J$ for two different times (the time $t$ is measured in units of mean inverse frequency).   A convergence towards 
the theoretical results of Eq. (\ref{current1}) is evident indicating that our system reached a NESS.


\newpage

\end{document}